# The nuclear quadrupole interaction at $^{111}$Cd and $^{181}$Ta sites in anatase and rutile $TiO_2$: A TDPAC study

Satyendra Kumar Das*, Sanjay Vishwanath Thakare[1] and Tilman Butz[2]

Radiochemistry Laboratory, Variable Energy Cyclotron Centre
Bhabha Atomic Research Centre, 1/AF Bidhan nagar, Kolkata 700064, India

[1]Radiopharmaceuticals Division, Bhabha Atomic Research Centre, Mumbai 400085, India

[2]Universität Leipzig, Fakultät für Physik und Geowissenschaften,
Institut für Experimentelle Physik II, Linnéstraße 5, 04103 Leipzig, Germany

The nuclear quadrupole interaction of the I=5/2 state of the nuclear probes $^{111}$Cd and $^{181}$Ta in the anatase and rutile polymorphs of bulk $TiO_2$ was studied using the time differential perturbed angular correlation (TDPAC). The fast-slow coincidence setup is based on the CAMAC electronics. For anatase, the asymmetry of the electric field gradient was η=0.22(1) and a quadrupole interaction frequency of $\omega_Q$=44.01(3) Mrad/s was obtained for $^{181}$Ta. For rutile, the respective values are η=0.56(1) and $\omega_Q$=130.07(9) Mrad/s. The values for rutile match closely with the literature values. In case of the $^{111}$Cd probe produced from the beta decay of $^{111}$Ag, the quadrupole interaction frequency and the asymmetry parameter for anatase was negligible. This indicates an unperturbed angular correlation in anatase. On the other hand for rutile, the quadrupole frequency is $\omega_Q$=61.74(2) Mrad/s and the asymmetry η=0.23(1) for $^{111}$Cd probe. The results have been interpreted in terms of the surrounding atom positions in the lattice and the charge state of the probe nucleus.

Keywords: TDPAC, anatase and rutile $TiO_2$, CAMAC, Nuclear quadrupole interaction.

PACS number: 61.66Fn, 07.05.Hd, 71.20.Nr

*Author for communication:
email: satyen@veccal.ernet.in

## Introduction

The interest in the study of $TiO_2$ matrix lies in the fact that in last few decades, $TiO_2$, a large band gap semiconductor [1] (3.2 eV for anatase and 3.06 eV for rutile) has emerged as one of the very important materials. It not only has developed interest in the basic research [2] but also has emerged as technologically important material [3]. $TiO_2$ either pure or in doped form has uses as photocatalyst for water purification, energy converter in solar cells, white pigment in paint, lipsticks, drugs, toothpaste, sunscreen material for the protection from UV light – to name a few. $TiO_2$ exists in different polymorphs of which the two important forms are anatase and rutile. Anatase is kinetically stable but rutile is thermodynamically stable at higher temperature. Anatase $TiO_2$ transforms irreversibly to rutile at higher temperature. These two modifications have uses depending on specific applications [4]. Both these polymorphs have tetragonal structures with different space groups, viz. $I4_1/amd$(141)for anatase and $P4_2/mnm$(136) for rutile. This structure indicates a symmetric EFG. For oxygen it is 2mm for both these modifications. However, a nonzero $\eta$ in both modifications has been observed. This needs clarification. TDPAC studies in case of anatase $TiO_2$ is scanty. Our earlier work [5] is one of the first TDPAC studies using $^{44}$Sc as the probe. In the present work we used two probes viz. $^{181}$Ta and $^{111}$Cd which have different valence states. Thus the chemistries of these probes with the host matrix are also different. The quadrupole interaction parameters can thus be good tools to understand the probe-host interaction. Another aspect concerning the dependence of the origin of the probe or decay mode through which the probe is formed has been addressed in this paper.

**Experimental:**

*Electronic circuit*

The electronics setup is comprised of three $BaF_2$ detectors (30mmx30mm cylindrical) coupled to XP2020Q photomultiplier tubes. DC coupled anode outputs were used for the timing purpose. Pulses from the tenth dynode and processed by the preamplifier have been used for the energy signal. CFD output for the start signal was directly fed into the TAC start input. However, the two stop signals (90° and 180°) from the CFDs were fed to the logic OR gate module (Philips model 755) through the gate and delay generators (GDG) which were used to change the delay and increase the pulse amplitude of the CFD outputs required by the logic unit.

Three amplifier outputs and the TAC output were the inputs for the CAMAC ADC (ORTEC model AD413A). The TAC SCA output was used as the master gate for the ADC. The crate controller (CC2000) [6] worked as the interface between the ADC and the PC which was used to provide instruction to the ADC through the software AMPS [7] used for the data acquisition, application of energy gates and data analysis. Data were recorded event by event in LIST mode with energy and time tagged on it. Coincidence events for the specific energy cascade were obtained by analysis of the LIST data by the software AMPS.

*Sample preparation: $TiO_2$ doped with $^{111}Ag$ and $^{181}Hf$*

The preparation of anatase $TiO_2$ and labeling with nuclear probe was done by the hydrolysis of Ti-tetra isopropoxide in presence of the probe. The details of this preparation is described elsewhere [5]. There is a small difference in chemistry for doping $^{111}Ag$ in $TiO_2$. The hydrolysis process in this case was carried out with NaOH solution instead of ammonia since silver dissolves in ammonia. $^{111}Ag$ probe was used in the carrier free form. This was obtained by chemical separation [8] from the Pd matrix which was used as target to produce $^{111}$Ag by neutron irradiation. The precipitates were thoroughly washed with water to remove any ammonia or NaOH present in the samples. They were then dried and annealed at different temperatures for nearly four hours in air. It was observed that $TiO_2$ annealed at 823K remained in the anatase phase and it was converted into rutile phase at 1123K. The samples annealed at different temperatures were counted by the coincidence setup mentioned above.

*TDPAC measurement:*

The time resolution of the coincidence setup was measured using $^{106}$Ru decaying to $^{106}$Rh which is in transient equilibrium with $^{106}$Ru. The latter undergoes beta decay and the cascade 622-512 keV comes from the deexcitation of the daughter $^{106}$Pd. The intermediate state of this cascade has a half life of 12 ps. Coincidence data were stored in LIST mode and the prompt spectra were obtained when the energy gates were applied on the Compton background depending on the cascade energy of the probe used in the experiment. In this case the cascade energies have to be below 512 keV. The prompt resolution (FWHM) measured for 97-245 keV($^{111}$Cd probe) was 1.25(2) ns and for 133-482 keV($^{181}$Ta probe) it was 1.11(5) ns.

Coincidences for 133-482 keV cascade of $^{181}$Hf and for 97-245 keV cascade of $^{111}$Ag were measured at 90° and 180°. Due to small abundances of the gamma lines of the $^{111}$Ag/$^{111}$Cd probe, the coincidence rate for the 97-245 keV cascade was very poor and the sample was to be counted for 3-4 days to achieve reasonable coincidence counts. The coincidences at 90° and 180° were used to obtain the perturbation factor $A_2G_2$ described below. The energy gates were adjusted in the software to get the best possible time spectrum with least prompt contribution at individual angle.

**TDPAC theory:**

The nuclear quadrapole interaction of the I=5/2 intermediate state leads to the splitting with three eigenvalues, $E_1$, $E_2$ and $E_3$ represented by [9]:

$$E_1 = -2rcos\left(\frac{\phi}{3}\right)$$

$$E_2 = rcos\left(\frac{\phi}{3}\right) - \sqrt{3}rsin\left(\frac{\phi}{3}\right)$$

$$E_3 = rcos\left(\frac{\phi}{3}\right) + \sqrt{3}rsin\left(\frac{\phi}{3}\right)$$

With

$$cos\phi = {}^{q}/_{r^3}$$

$$r = sign(q)\sqrt{|p|}$$

$$p = -28\left(1 + {}^{\eta^2}/_{3}\right)$$

$$q = -80(1 - \eta^2)$$

Where η is the asymmetry of the electric field gradient mentioned below.

The three precession frequencies are:

$$\omega_1 = (E_2 - E_3)\omega_Q$$

$$\omega_2 = (E_1 - E_2)\omega_Q$$

$$\omega_3 = (E_1 - E_3)\omega_Q$$

With quadrupole frequency is:

$$\omega_Q = \frac{eQV_{zz}}{40\hbar}$$

where Q denotes the quadrapole moment of the I = 5/2 state and $V_{zz}$ is the largest component in magnitude of the EFG tensor. The perturbation function is thus:

$$G_2(t) = a_0 + a_1 cos\omega_1 t + a_2 cos\omega_2 t + a_3 cos\omega_3 t$$

Experimental data were fitted with this function modified with the finite distribution of $\omega_Q$. In the present case we have used lorentzian distribution.

TDPAC spectra for the anatase and rutile $TiO_2$ with $^{181}$Ta probe are shown in figure-1. On the left of the figure are shown the $A_2G_2$ spectra and the corresponding cosine transforms are shown on the right. Figure-2 represents the results when $^{111}$Ag/$^{111}$Cd probe was used. The fitted results along with the literature data are shown in table-1.

**Discussion**

In this present work we address some aspects related to fast-slow coincidence circuit based on the CAMAC electronics adopted for TDPAC studies and the results of nuclear quadrupole interaction in anatase and rutile $TiO_2$ with different probes. Performance of this system was checked by measuring the quadrupole interaction parameters for known systems viz. rutile $TiO_2$ doped with $^{181}$Hf/$^{181}$Ta probe and $AgNO_3$ crystals doped with $^{111}$Ag/$^{111}$Cd probe. The quadrupole frequency and the asymmetry of EFG were found to match closely with literature values for both the systems [10,11]. Rutile TiO2 was prepared by us [5] through a different method than the chemical route followed by others [11]. Close proximity of the interaction parameters indicate that the rutile TiO2 prepared through different chemical methods are same. The results of $TiO_2$ for both $^{181}$Ta and $^{111}$Cd probes along with other literature data are shown in table-2.

The result for the anatase $TiO_2$ presented in this work has been obtained for the first time by us. In our previous work [5], we found that wide variation in the values of Vzz and $\eta$ exists depending on the probe. Any probe other than the indigenous atom Ti is an impurity in the host matrix. The chemical bonding of this impurity with the host matrix is different than in the indigenous matrix – which leads to the difference in the hyperfine parameters for different probes. The crystal symmetry in anatase is 4m2 which implies the axial symmetry ($\eta$ = 0). However a nonzero $\eta$ (=0.22) for the probe $^{181}$Ta can be attributed partly due to the relatively larger distribution width of $\omega_Q$. Poor crystallinity due to the low temperature annealing was attributed to this observation. In another work we adsorbed hafnium ion from aqueous solution on commercially available $TiO_2$ and it was annealed at 1123K. Preliminary results indicate the presence of both anatase and rutile. The width of the frequency distribution for the anatse phase in this sample also was relatively large and it could not be improved to the extent of rutile. It is not thus possible at this stage to assign the definite reason for the relatively broad distribution in anatase phase. Another reason for the nonzero $\eta$ could be the 0.1% impurity due to the probe nucleus in the $TiO_2$ matrix.

For the $^{111}$Ag/$^{111}$Cd probe, the flat response of the perturbation function and also the FFT without any line indicates that there is no perturbation of in the anatase phase. Unfortunately there is no literature data with $^{111}$In/$^{111}$Cd probe in the same system. However in the case of rutile phase, the results in table-2 indicate a strong dependence of the parent giving rise to the same probe ($^{111}$Cd). When the parent is a $\beta$- decaying nucleus ($^{111}$Ag), the EFG is almost three times larger compared to that for the EC decaying parent ($^{111}$In). A plausible explanation is given below.

Table-2 indicates that for both probes, $V_{zz}$ for rutile is more than in anatase. This can be explained in terms of the O-positions in the lattice around the probe nucleus. There are two sets of O- atoms around the probe atom: four O- neighbours (with M-O distance, say, d1)

and two O- neighbours (with M-O distance, say, d2) in the $MO_6$ octahedra. The calculated values of d1 and d2 in anatase and rutile phases vary depending on the probe and the number of electrons to be added over the indigenous Ti ion. It is found [12] that though d1 increases, there is decrease in d2 in rutile compared to those in anatase phase. Using these M-O parameters, Vzz in rutile has been found to be more in rutile for both probes $^{111}$Cd [5, 12] and $^{181}$Ta [11]. This, however, does not explain why $V_{zz}$'s are different for the same probe originating from different parents. Is the after-effect of the EC decay which is responsible for the lower value of the $V_{zz}$ in $^{111}$In-$^{111}$Cd probe? We have not observed any indication of time dependent interaction in the TDPAC spectrum. It is known that the EC decay is an energetic process which causes lot of shake off in the electron shells leaving the daughter probe in high charged state. In other words, there is deficiency of electron density around the probe nucleus in the EC decay process. If the relaxation is not very fast, it is possible that the $V_{zz}$ will be less due to low electron density in the case of EC decay compared to that in beta decaying parent. Theoretical calculation is underway to explain the details of bonding of the probe atoms and the role of the charge states of the daughter due to difference in the decay modes of the parents in these two polymorphs of $TiO_2$.

**Conclusion**

Bulk $TiO_2$ labeled with $^{111}$Ag-$^{111}$Cd and $^{181}$Hf-$^{181}$Ta probes have been studied by TDPAC technique. A new methodology of coincidence counting using CAMAC electronics has been introduced. A general trend of higher $V_{zz}$ in the rutile phase than in anatase phase has been observed in this work. This is true for other probes as well and was explained in terms of the O-positions around the probe nucleus. Further, the same probe but with different parents show a very different trend. $^{111}$Cd from different parents viz. $^{111}$Ag and $^{111}$In show very different $V_{zz}$ particularly in the rutile phase. This has been explained on the basis of charge states of the daughter from different nuclear decay processes from different parents.

Nano $TiO_2$ and the thin film of $TiO_2$ are important materials for several applications and understanding the structures of these materials would be interesting from basic research point of view. The results for the bulk $TiO_2$ described in this work would be a guideline to interpret the data for these materials to understand the structural aspects. Research is in progress in this direction.

**Acknowledgement**

Authors are thankful to Dr. S.K. Bandopadhyay, Material Science Group, Variable Energy Cyclotron Centre (VECC) for his useful comments. They are also thankful to Mr. R.K. Chatterjee, Radiochemistry Laboratory, VECC, for his help in preparing the samples and taking data.

**References:**

1. T. Toyoda, H. Kawano, Q. Shen, A. Kotera and M. Ohmori, Jpn. J. Phys. Part-I 39. 3160 (2000).
2. Zhenrong Zhang, Quingfeng Ge, Shao-chun Li, Bruce D. Kay, J.M. White, and Zdenek Dohnálek, Phys. Rev. Lett. 99, 126105(2007).
3. Yun Gao, Saipeng Wong, Quan Li, Kai Hong Cheng, Ning Ke, Wing Yin Cheung and Guo Sheng Shao, Japanese Journal of Applied Physics 46(9A) 5767(2007).
4. Amy A. Gibb and Jullian F. Banfield, American Minerologist 82, 717(1997).
5. Seung-Baek Ryu, Satyendra Kumar Das, Tilman Butz, Werner Schmitz, Christian Spiel,, Peter Blaha, and Karlheinz Schwarz, Phys. Rev. B77, 094124(2008).
6. S.P. Borkar et. al. Symposium on Advanced Instrumentation for Nuclear Research, (SAINR)-93, BARC, Trombay, India.
7. A. Chatterjee, Sushil Kamerkar, A.K. Jethra, S. Padmini, M.P. Diwakar, S.S. Pande and M. D. Ghodgaonkar, Pramana, 57(1), 135(2001).

8. K.V. Vimalnath, S. Saha, V. Chirayil, Proc. DAE-BRNS Symp. On Nuclear and Radiochemistry, NUCAR-2007, 14-17 Feb. 2007 India, pp. 559-560.

9. T. Butz, Hyp. Int. 52, 189(1989), Erratum: T. Butz, Hyp. Int. 73, 387(1992).

10. H. Haas and D.A. Shirley, J. Chem. Phys. 58(8), 3339(1973).

11. James A. Adams and Gary L. Catchen, Phys. Rev. B50(2), 1264(1994).

12. L.A. Errico, G. Fabricius and M. Renteria, Phys. Rev. B67 144104 (2003).

Table-1: TDPAC parameters for anatase and rutile $TiO_2$ for different probes

| Probe | $TiO_2$(anatase) | | $TiO_2$(rutile) | |
|---|---|---|---|---|
| | $\omega_Q$ (Mrad/s) | $\eta$ | $\omega_Q$ (Mrad/s) | $\eta$ |
| $^{111}Ag/^{111}Cd$ | ~0.0 | -- | 61.74(2) | 0.23(1) |
| $^{181}Hf/^{181}Ta$ | 44.01(3) | 0.22(1) | 130.07(9) | 0.56(1) |

Table-2: Hyperfine parameters for anatase and rutile $TiO_2$ with different probes.

| Probe | Anatase | | Rutile | | Remarks |
|---|---|---|---|---|---|
| | $V_{zz}$ $x10^{17}Vcm^{-2}$ | $\eta$ | $V_{zz}$ $x10^{17}Vcm^{-2}$ | $\eta$ | |
| $^{111}Ag/^{111}Cd$ | ~0.0 | _ | 20.3(2) | 0.23(1) | This work |
| $^{111}In/^{111}Cd$ | _ | _ | 5.80(2) | 0.22(2) | Ref[11] |
| $^{181}Hf/^{181}Ta$ | 4.62(1) | 0.22(1) | 13.65(10) | 0.56(1) | This work |
| $^{181}Hf/^{181}Ta$ | _ | _ | 13.30(6) | 0.57(1) | Ref[11] |

**Figure caption**:

Fig-1: TDPAC spectra with cosine transform for anatase and rutile $TiO_2$ labeled with $^{181}Ta$ probe.

Fig-2: TDPAC spectra with cosine transform for anatase and rutile $TiO_2$ labeled with $^{111}Cd$ probe.

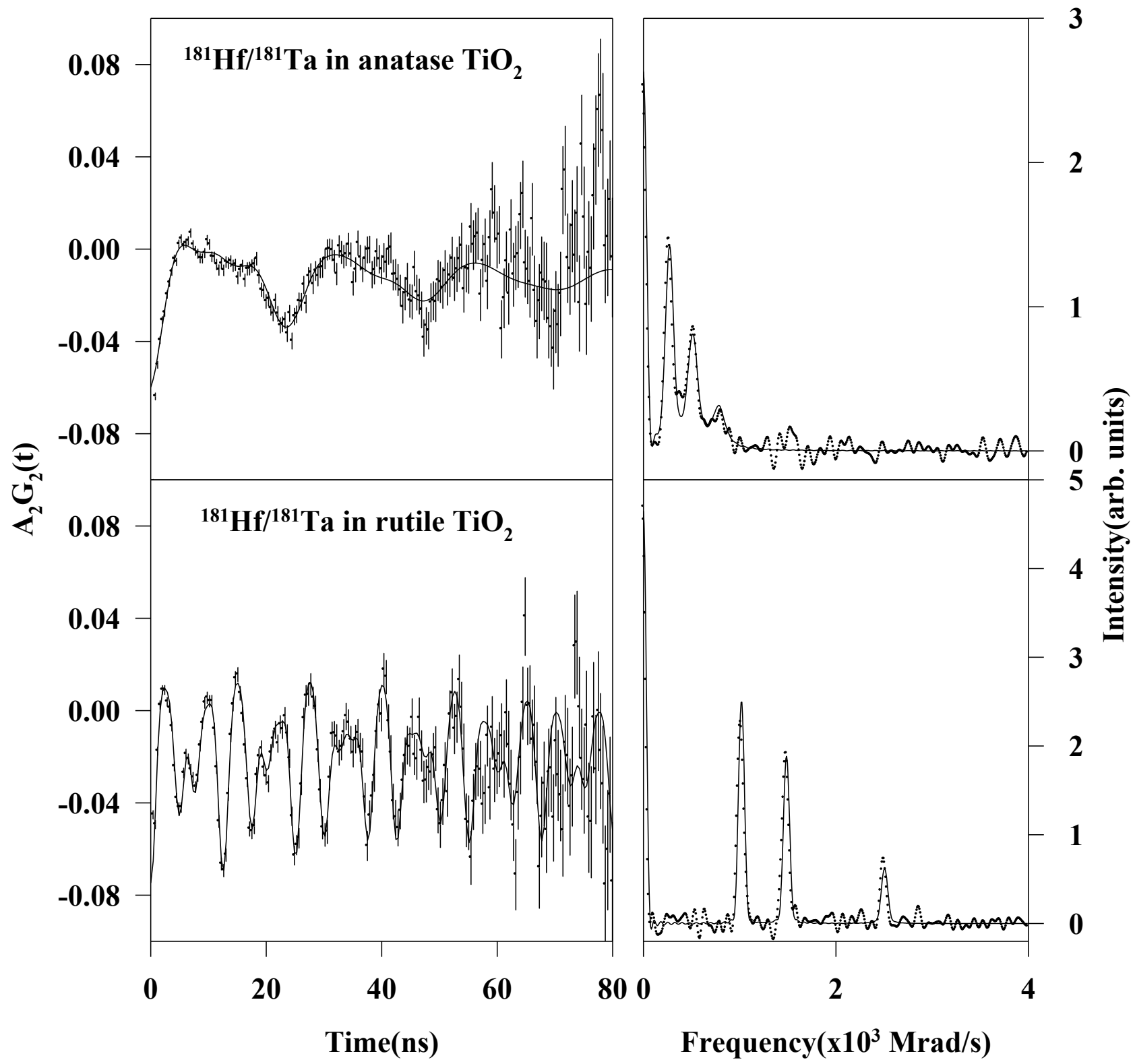

181Hf/181Ta in anatase TiO2
181Hf/181Ta in rutile TiO2
A2G2(t)
0.08
0.04
0.00
-0.04
-0.08
0
20
40
60
80
Time(ns)
0
2
4
Frequency(x10³ Mrad/s)
Intensity(arb. units)
3
2
1
0
5
4
3
2
1
0

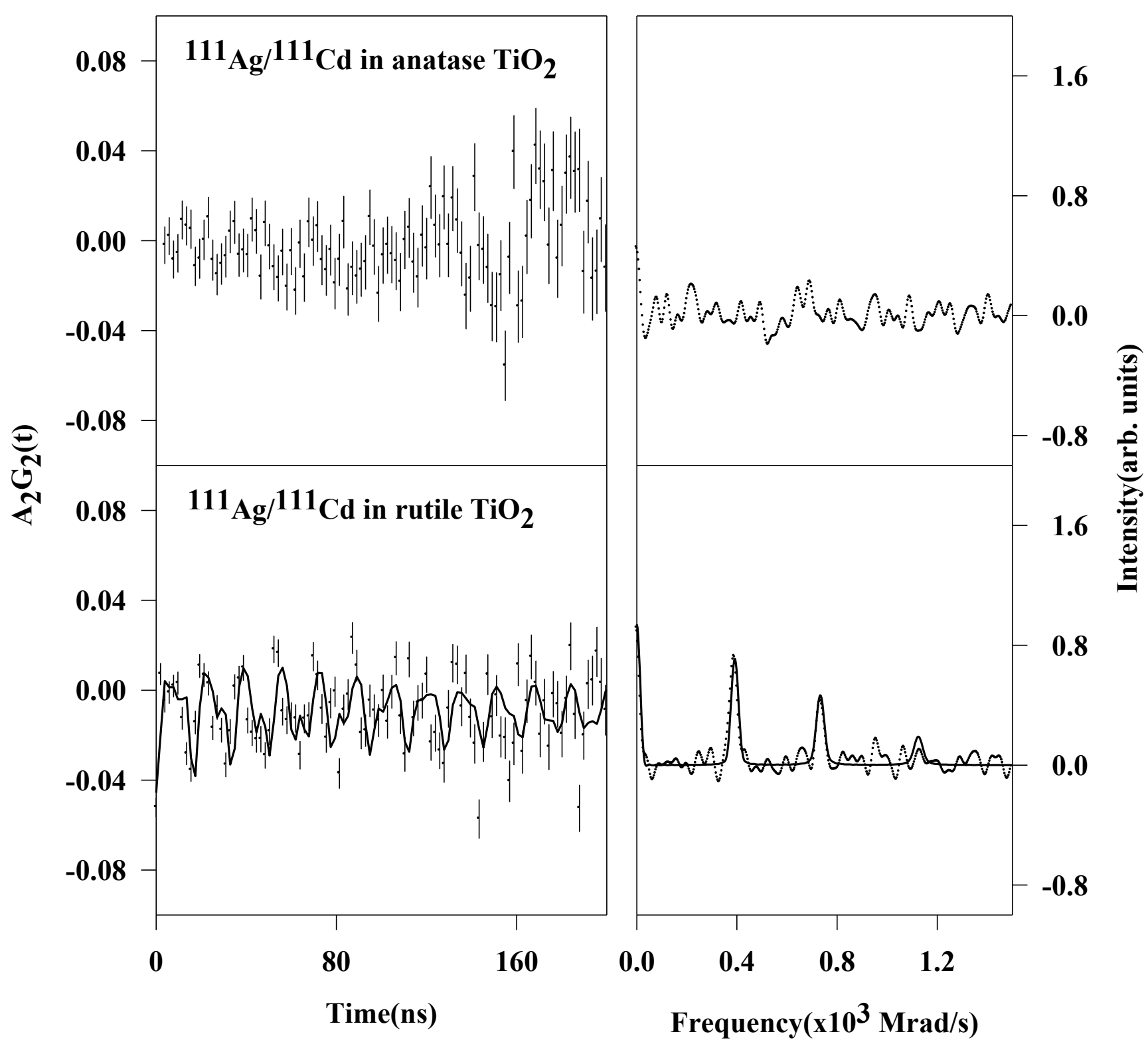

$^{111}Ag/^{111}Cd$ in anatase $TiO_2$
$^{111}Ag/^{111}Cd$ in rutile $TiO_2$
$A_2G_2(t)$
0.08
0.04
0.00
-0.04
-0.08
0
80
160
Time(ns)
1.6
0.8
0.0
-0.8
0.0
0.4
0.8
1.2
Frequency(x$10^3$ Mrad/s)
Intensity(arb. units)